\documentclass[%
 reprint,
 amsmath,amssymb,
 aps,
]{revtex4-2}

\usepackage{graphicx}
\usepackage{dcolumn}
\usepackage{bm}
\usepackage{color}

\begin{document}

\preprint{APS/123-QED}

\title{pH-Dependent Zeta Potential \\ Induces Diffusiophoretic Focusing  in an Acid-Base Reaction}

\author{Ethan Coleman}
\author{Ankur Gupta}%
\email{ankur.gupta@colorado.edu}
\affiliation{%
 Department of Chemical and Biological Engineering, University of Colorado, Boulder, Colorado 80309, USA 
}%

\date{\today}

\begin{abstract}
    Diffusiophoresis of charged particles in the presence of electrolytes has been extensively studied in the literature. However, in these setups, particles typically move in a single direction, either up or down the electrolyte gradient. Here, we theoretically investigate the conditions under which a particle can reverse its diffusiophoretic direction within the same setup, leading to the formation of a focusing band under steady-state concentration gradients. Using multi-ion diffusiophoresis calculations, we simulate particle transport in an acid-based reaction system where salt is added alongside the acid. For a range of salt concentrations, particles focus within the channel. Our analysis reveals that a pH-dependent zeta potential is necessary for this focusing to occur, and determines where the particles focus, i.e., on or off the acid-base reaction front. We report qualitative agreement with prior experimental observations and derive analytical conditions governing particle focusing, highlighting the delicate balance between concentration gradients and zeta potential variations. The work elucidates the crucial physics of pH-dependent zeta potential and opens new avenues for exploring diffusiophoresis in acid-base systems, with implications for microfluidic design and biophysical transport processes.
\end{abstract}

\maketitle


Diffusiophoresis refers to the advective transport of microparticles in response to the gradients of chemical species. Over the past two decades, there has been a significant interest in employing diffusiophoresis to create living crystals \cite{crystals}, assemble self-spinning microgears \cite{gear},  understand biological pattern formation \cite{turing},  laundry action \cite{surfactant}, membraneless water filtration \cite{membraneless} and enhancing transport of colloids \cite{boost}, among others.  We focus on electrolytic diffusiophoresis, where an electrolyte gradient transports charged particles via electrophoresis and chemiphoresis~\cite{ec1,ec2,ec3,ec4,ec5,ec6,ec7,ec8,ec9,ec10,ec11}. Often, electrolytic diffusiophoresis is studied in dead-end pores \cite{dep1,dep2,dep3,dep4,dep5,dep6,dep7,dep8,dep9,dep10,dep11}, T-junctions \cite{tj1,tj2}, microchannels \cite{boost,mc1,mc2, trap1,trap2},  and 2-dimensional and porous geometries \cite{bannarjee,por1,por2,por3,por4,por5,por6, sambamoorthy2025diffusiophoresis, sambamoorthy2023diffusiophoresis}.  

\begin{figure}
\includegraphics[width=\linewidth]{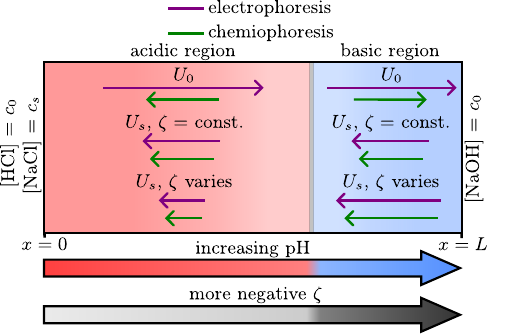}
\caption{\label{fig:schematic} Overview of the setup and qualitative description of relevant physics. HCl and NaCl are present at a fixed concentration of [HCl]=$c_0=0.5$ mM and [NaCl]=$c_s$ at $x=0$. Similarly,  NaOH is present at a fixed concentration of [NaOH]=$c_0=0.5$ mM at $x=L$. A steady-state concentration gradient exists, with pH increasing from left to right, and a sharp change occurs across the acid-base reaction front at $x = x_r$. We refer to the region $x<x_r$ as the acidic side and the region $x>x_r$ as the basic side.  At $t=0^+$, a uniform concentration of polystyrene particles is introduced which move diffusiophoretically in response to the concentration gradients of the electrolytes.  For qualitative description, we decompose the net diffusiophoretic velocity $U$ as the superposition of $U_0$ and $U_s$, where $U_0$ is the velocity due to HCl and NaOH only, or $c_s=0$ and $c_0 \neq 0$, and $U_s$ is the change in velocity induced by adding salt $c_s \neq 0$. For focusing to occur, $U(x<x_r)>0$ and $U (x>x_r) <0$. For the setup of interest, $U_0>0$ for all $x$ and is higher in the basic region than the acidic region due to cooperation between the elecrophoresis and chemiphoresis contributions. $U_s <0$ and is approximately constant if one assumed a constant $\zeta$ potential. However, if we assume $\zeta$ potential to a function of pH, where zeta potential is higher in the basic side than the acidic side, $U_s$ becomes strongly leftwards in the basic region, potentially causing the particles to move leftwards in the basic region but rightwards in the acidic region, and thus inducing focusing. }
\end{figure}

Even though electrolytic diffusiophoresis has been widely studied, the particles tend to simply move towards or against the region of higher electrolyte concentration, depending on their mobility (although diffusioosmosis can reverse this trend in some specific cases~\cite{tj2}). This invites a fundamental question: can we design systems where particles can switch their direction within the same experiment, even at steady state? To the best of our knowledge, the only study that reports such a phenomenon is Shi. et al~\cite{main}, where the authors report focusing of polystyrene particles due to competing pH and NaCl gradients. The setup, see Fig.~\ref{fig:schematic}, consists of HCl on the left-hand side $\left(x=0\right)$ with [HCl]=$c_0$ and [NaOH] on the right-hand side $\left(x=L\right)$ with [NaOH]=$c_0$, which creates an acid-base reaction front at $x=x_r$. A salt gradient is established by NaCl, with [NaCl]=$c_s$ at $x=0$. We assume that uniformly distributed polystyrene particles respond to the concentration gradients setup by the electrolytes at $x=0$ and $x=L$. Particle focusing can occur for some values of $c_s$ where the diffusiophoretically induced velocity, $U$, switches sign from positive to negative as we go from left to right and the particles focus at $x=x_f$ where the velocity is zero. For focusing to occur, two conditions must be satisfied: $U (x<x_f) > 0$, and $U(x>x_f) < 0$. Shi et al. ~\cite{main} proposed a simplified condition $U (x<x_r) < U (x>x_r)$, and while this restricts the choice of ions that can lead to focusing, it does not conclusively predict when focusing occurs. It also assumes $x_f = x_r$, which is not always true, losing information about where focusing occurs. In this Letter, we underscore pH-dependent zeta potential -- a factor commonly ignored in diffusiophoresis literature ~\cite{ec7,main,moller2017steep,seo2020continuous,lee2018diffusiophoretic} -- is required for particles to focus. We derive strict semi-analytic conditions capable of predicting ranges of $c_s$ that induce focusing, as well as where in the channel this focusing occurs. This new understanding addresses the question: when, why, and where the diffusiophoretic particles focus.


Before diving into the mathematical details, we explain the essential physics required for focusing by providing a qualitative description of the system dynamics. This discussion is primarily aimed to provide a physical intuition as to \textit{why} particles focus. We decompose $U = U_0 + U_s $, where $U_0$ is the velocity induced by the acid and base in the absence of salt, $c_s = 0$, and $U_s$ is the velocity induced by the addition of salt to the channel.  We also assume that particles focus at the reaction front; an assumption we relax in the full model. In reality, all the ions will interact with each other to set $U$~\cite{ankur, ec6}, but certain aspects of the relationship between $U_0$ and $U_s$ are known without a fully calculated solution. 

\par{} In this simplified picture, it is convenient to divide the channel into two parts: the acidic side ($x<x_r$) and the basic side ($x>x_r$). For focusing to occur, we require that particles move rightwards within the acidic side and move leftwards within the basic side, or $U(x<x_r)>0$ and $U(x>x_r)<0$. First, we focus on $U_0$, which is the particles' motion only due to the HCl and NaOH gradients. Polystyrene particles, which are negatively charged,  migrate rightwards -- or towards higher pH -- in both the acidic and basic regions. However, the particles inside the basic side move faster, as also observed in experiments \cite{main}. This happens because electrophoresis and chemiphoresis support each other in the basic region but oppose each other in the acidic gradients; see Fig.~\ref{fig:schematic}.
\par{} Since $U_0$ always transports particles rightwards, to induce focusing, $U_s$ should reverse the direction in the basic side, but not in the acidic side. Mathematically, this implies that $ -U_s(x>x_r) > U_0(x>x_r)$ and $-U_s(x<x_r) < U_0(x<x_r)$. 
\par{} Polystyrene particles diffusiophoretically transport towards higher concentrations of NaCl ~\cite{florea2014long, dep4, main, dep9, dep10, ankur, dep2}. Therefore, $U_s < 0$ throughout the channel. First, we focus on the scenario of constant zeta potential $\zeta$. Since $U_s$ is proportional to the log of NaCl concentration across the channel, for a constant $\zeta$, $U_s$ is roughly constant throughout the channel ($U_s$ does increase modestly in the basic side, but for a qualitative discussion, this can be ignored). Consequently, NaCl will decrease $U$ uniformly throughout the channel. For a sufficiently high magnitude zeta potential $\left| \zeta \right|$, $U_s$ will overcome $U_0$ and $U$ will become negative in some regions. However, since the acidic side possesses a smaller $U_0$, it would be the region to reverse velocity directions first. Mathematically, the condition that will be reached first is $U(x<x_r)<0$ and $U(x>x_r)>0$,  which implies particles \textit{defocus} and create a region with no particles. For focusing to occur, the basic side must reverse direction first, which is only possible if $|\zeta|$ increases with pH. 
\par{} Importantly, PS particles have a zeta potential that depends on pH \cite{main,ec7,timmerhuis2022diffusiophoretic}. We sketch qualitative pH values in Fig.~\ref{fig:schematic} where pH experiences a sudden jump at the reaction front, similar to what is observed in a titration curve.
\par{} Because the pH is relatively constant in each regime, the zeta potential experiences a similar jump, causing the basic region to have a larger $|\zeta|$, see Fig.~\ref{fig:schematic}.  The electrophoretic and chemiphoretic contributions are approximately proportional to $|\zeta|$ and $|\zeta|^2$~\cite{ec4, ec6, ankur}. Thus even a modest change in $\zeta$ can result in a high sensitivity to $U_s$. This increased sensitivity in the basic region leads to $-U_s (x>x_r) > U_0 (x>x_r)$ and $-U_s(x>x_r) < U_0(x<x_r)$, for certain values of $c_s$, making focusing possible. In summary, a pH-dependent zeta potential explains \textit{why} particles focus.
\par{} Next, we focus on a quantitative model to predict ion and particle transport. We write the Nernst-Planck equations at steady state 
\begin{subequations}
\label{eq:goveqn}
\begin{eqnarray}
    \frac{d N_i} {d x} = R_i \\
   N_i = - D_i \left(  \frac{d c_i}{d x}  - \frac{z_ie}{k_BT} c_i E \right), 
\end{eqnarray}
where $c_i$, $D_i$, $z_i$, $N_i$ and $R_i$  are the concentration, diffusion coefficient, valency, flux, and reaction consumption for the $i^{\textrm{th}}$ ion, Na$^+$, Cl$^-$, H$^+$, or OH$^-$. In addition, $e$ is the charge on an electron, $k_B$ is the Boltzmann constant, and $E$ is the electric field. For the described system, $R_i$ corresponds to acid-base reaction and $R_H = R_{OH} = -kc_Hc_{OH}$, where $k$ is a rate constant, while $R_{Na}=R_{Cl}=0$. Electorneutrality $\sum z_i c_i=0$ and the zero-current condition $\sum z_i N_i=0$ yield~\cite{gupta2019diffusion} 
\begin{equation}
    E = \frac{k_B T}{e}\frac{\sum_{i} D_i z_i \frac{d c_i}{d x}}{\sum_{i} D_i z_i^2 c_i}.
\end{equation}
\noindent For the pH-dependence of $\zeta$ of interest to this work, the relationship is approximately linear with $\zeta(\textrm{pH}_{acid} =4) = \zeta_{acid} \approx -49$~mV and $\zeta(\textrm{pH}_{base}=10) = \zeta_{base} \approx -61$~mV \cite{main, ec7}; see \cite{SI} for details. We simply write
\begin{equation}
\zeta(\textrm{pH}) \approx   - \frac{ \Delta \zeta }{\Delta \textrm{pH}} \left( \textrm{pH} - \textrm{pH}_{acid} \right) + \zeta_{acid},
\label{eq:ph}
\end{equation}
where $\Delta \zeta = \left| \zeta_{base} - \zeta_{acid} \right|$ and $\Delta \textrm{pH} = \textrm{pH}_{base} - \textrm{pH}_{acid}$. It is possible to use a physical approach to capture the dependence of zeta potential~\cite{ninham1971electrostatic} on pH instead of Eq.~\eqref{eq:ph} but we do not explore it in detail since the pH is roughly constant within the acidic and basic sides of the channels, making a simple model sufficient to capture the revelevant physics. {We do acknowledge, however, that these results are sensitive to pH, and will not hold for particles that experience non-linear pH dependence \cite{SI}.} By solving Eq.~\eqref{eq:goveqn} to obtain $c_i(x)$ at steady state, it is straightforward to evaluate $U(x)$ as~\cite{ankur}
\begin{equation}
\label{eq:multiU}
    U = \frac{\varepsilon}{\mu} \left(  E \zeta + \frac{\partial \log I}{\partial x} \frac{\zeta^2}{8} \right),
\end{equation}
\end{subequations}
where $\varepsilon$ and $\mu$ are the electrical permittivity and viscosity of the electrolyte solution, and $I= \frac{1}{2}\sum_i z_i^2 c_i$ is the ionic strength. We note that Eq.~\eqref{eq:multiU} is only valid in the low potential limit limit~\cite{ankur}. However,  the accuracy of this expression is reasonable for $\tilde{\zeta} = \left| \frac{e \zeta}{k_BT} \right| \lesssim 4$~\cite{ankur}. 
\begin{figure}
\includegraphics[width=0.48\textwidth]{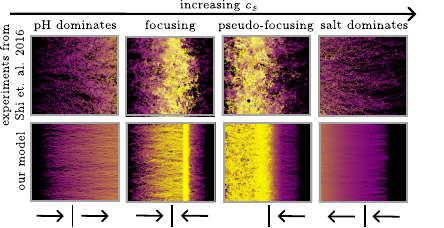}
\caption{\label{fig:exp} Comparison of model results with experimental observations from Shi et al.~\cite{main}; experimental results are adapted with permission. Concentration profiles of ions are calculated by solving Eq.~\eqref{eq:goveqn} and particle trajectories are visualized via the Langevin equation. The particle streak lines represent 0 seconds (purple) to 60 seconds (yellow). The model can predict all the regimes: pH-dominated, focusing, pseudo-focusing, and salt-dominated. The arrows represent the direction of $U$. $c_0=0.5$ mM across all cases. Experiments represent $c_s=[0, 0.5, 1, 2]$ mM from left to right whereas the model utilizes  $c_s=[2.5, 3, 3.5, 4]$ mM. The reasons for quantitative discrepancies between $c_s$ values in the experiments and the model are discussed in the text.}
\end{figure}
Finally, we calculate particle trajectories using the Langevin equation; see details in \cite{SI}. We note that similar to Shi et al.~\cite{main}, we focus on the scenario when the concentration is at a steady state and the particles respond to the steady gradients. A summary of the four different scenarios from particle simulations and their comparison with experiments from Shi et al.~\cite{main} is outlined in Fig.~\ref{fig:exp}. \par{}
We plot the trajectories until $t=60$ seconds and observe compelling qualitative agreement with the prior experimental results. While the experiment observed focusing with a salt concentration of $c_s = 0.5$  mM, our model focuses for $c_s = 3$  mM. However, step sizes between cases reported are the same, so the focusing window follows the same trend; the model is simply shifted by $2.5$ mM compared to the experiments. This result is self-consistent with our model, since we derive the necessary condition for focusing that the flux of Na$^+$ must be greater than zero, which occurs only for $c_s > c_0$ \cite{SI}. However, we acknowledge that our low potential assumption may artificially lead to decreased sensitivity, which could be the cause of this discrepancy. Even though the comparison is qualitative, we recover all the essential features in experiments. 
\begin{figure}
\centering
\includegraphics[width=\linewidth]{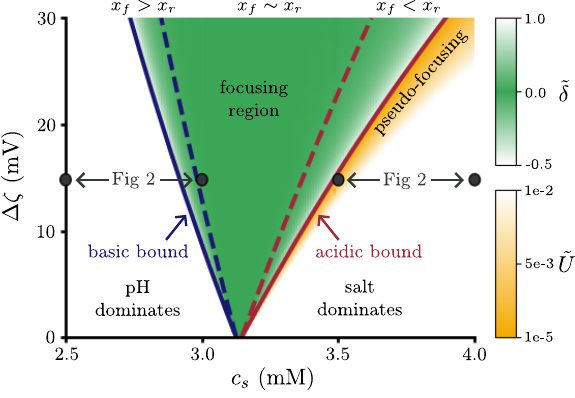}
\caption{Phase diagram of different regimes as a function of $\Delta \zeta$ and $c_s$. We vary $\Delta \zeta$ such that $\zeta \left(x=0, \textrm{pH}=3.3 \right) = -55 + \frac{\Delta \zeta}{2}$ mV and $\zeta \left(x=1, \textrm{pH}=10.7 \right) = -55  - \frac{\Delta \zeta}{2}$ mV. The velocity $U$ is calculated by solving Eq.~\eqref{eq:goveqn} and if $U$ changes from positive to negative, we refer to that as the focusing regime. We calculate $\tilde{\delta} = 1 - \frac{x_f}{x_r}$ to differentiate between on-front and off-front focusing. Finally, we define $\tilde{U} = \frac{U_{\textrm{min}}}{U_{\textrm{max}}}$ to quantify the psuedofocusing regime if $\tilde{U} < 10^{-2}$.  The dashed and solid lines are predictions from Eq.~\eqref{eq:focus}. The conditions of Fig. 2 are highlighted in black circles.}
\label{fig:focus}
\end{figure}  Before focusing on the location of focusing $x_f$, we first describe the two limits without focusing. For $c_s \ll c_0$, the salt concentration is minimal and hence diffusiophoresis is driven by gradients HCl and NaOH, both of which cause the particles to move rightwards; see Fig.~\ref{fig:exp}. Similarly, when $c_s \gg c_0$, the NaCl dominates and particles move leftwards. Therefore, focusing occurs between these two limits.
\par{} We recall that $x_f$ is the location where $U=0$. Note that $U$ is sensitive to both $c_i$  and $\zeta$; see Eq.~\eqref{eq:multiU}.  However, the sensitivity is highest with $\zeta$. Therefore, the most likely location for the particles to focus is the reaction front where $\zeta$ jumps abruptly by $\sim 10$ mV, or $x_f \approx x_r$; see Fig.~\ref{fig:exp}. This is consistent with our numerical simulations. However, it is certainly possible to have off-front focusing, as $\zeta$ varies by $\sim 2.5$ mV in each region. We find that in addition to off-front focusing, the acidic region experiences cases where the velocity vanishes $U(x < x_r) \approx 0$, but never truly switches sign~\cite{SI}. We refer to this effect as pseudo-focusing, noting that in this scenario, the velocity magnitude is small, and hence particle migration is minimal, creating a wider focusing region; see Fig.~\ref{fig:exp}. 
\par{} The above discussion highlights that focusing occurs when the addition of salt reverses diffusiophoretic velocity in the basic region but not in the acidic region. This reversal happens when zeta potential is pH-dependent. Finally, the focusing can occur both at the reaction front or off the reaction front, and can also be achieved via pseudo-focusing. \par{} We perform simulations by varying $\Delta \zeta$ and $c_s$ to create a phase diagram of different focusing regimes. The results are outlined in Fig.~\ref{fig:focus}.  We assume $\zeta \left(\textrm{pH}=7 \right)=-55$ mV and $c_0=0.5$ mM remain constant. We vary $c_s$ and $\Delta \zeta$ such that $\zeta \left(x=0, \textrm{pH}=3.3 \right) = -55 + \frac{\Delta \zeta}{2}$ mV and $\zeta \left(x=L, \textrm{pH}=10.7 \right) = -55  - \frac{\Delta \zeta}{2}$ mV; note $\Delta \zeta > 0$. The variation of the zeta potential with pH is considered to be linear, as shown in Eq.~\eqref{eq:ph}. We employ Eq.~\eqref{eq:goveqn} to calculate $N_H$, $N_{OH}$, $N_{Cl}$ and $N_{Na}$ for different $c_s$ values, and consequently map out the ion concentration profiles \cite{SI} and $U$ as a function of $\Delta \zeta$ and $c_s$. If $U$ switches direction from positive to negative as we move from left to right, we refer to it as focusing. We define $\tilde{\delta} = 1 - x_f/x_r$ such that when $\delta=0$, we observe on-front focusing and when $\delta \neq 0$, we observe off-front focusing. In addition, we define $\tilde{U} = U_{min}/U_{max}$ such that if $\tilde{U}<10^{-2}$, we refer to the condition as pseudo-focusing~\cite{SI}. Fig.~\ref{fig:focus} highlights that pH-dependent zeta potential is practically required to induce focusing.  Furthermore, a higher pH-dependency on zeta potential expands the range of salt concentrations for which focusing can occur.
\par{} To better understand the conditions that lead to focusing and to make analytical progress, we analyze the system under steady-state conditions while also assuming an instantaneous acid-base reaction. This allows us to approach the problem as two separate regions, $x<x_r$ and $x>x_r$. We denote the concentrations of each ionic species as $c_i^{a}$ and $c_i^{b}$ for the acidic and basic sides, respectively. This enables us to argue that $c_{OH}^{a}=0$ and $c_{H}^{b}$=0, and the problem reduces to a three-ion analysis in each region. After invoking electroneutrality and zero-current conditions, the solution in the acidic and basic side become~\cite{SI}
\begin{subequations}
\label{eq:3ion}
\begin{widetext}
\begin{eqnarray}
U^{a} = -\frac{\varepsilon \zeta}{\mu c_{Cl}^a} \left[ \frac{N_H}{D_{HCl}} \left( \frac{k_B T}{e}  \beta_{HCl}  + \frac{\zeta}{8}  \right) +  
 \frac{N_{Na}}{D_{NaCl}} \left( \frac{k_B T}{e} \beta_{NaCl} + \frac{{\zeta}}{8}  \right) \right], \label{eq:3ion_acid}\\
U^{b} = -\frac{\varepsilon \zeta}{\mu c_{Na}^b} \left[ \frac{N_{OH}}{D_{NaOH} } \left( \frac{k_B T}{e} \beta_{NaOH}  + \frac{\zeta}{8}  \right) +  
 \frac{N_{Cl}}{D_{NaCl} } \left( \frac{k_B T}{e} \beta_{NaCl} + \frac{{\zeta}}{8}  \right) \right], \label{eq:3ion_base}
\end{eqnarray}
\end{widetext}
\end{subequations}
where $D_{HCl}$,  $D_{NaCl}$ and $D_{NaOH}$ are the ambipolar diffusivities of the three electrolytes~\cite{deen}. Similarly, $\beta_{HCl}$, $\beta_{NaCl}$, and $\beta_{NaOH}$ are the diffusivity contrast ratios, such as $\beta_{HCl} = \frac{ D_H - D_{Cl}}{D_H + D_{Cl}}$. 
\par{} Eq.~\eqref{eq:3ion} offers expression for the mobility of a three-ion system and it is an elegant extension to the mobility of a binary electrolyte. In Eq.~\eqref{eq:3ion_acid}, the first term within the bracket of the right-hand side refers to the contribution to $U^a$ due to the electrolyte HCl. Similarly, the second term on the right-hand side captures the contribution due to NaCl. Since Cl$^-$ is the common ion between HCl and NaCl, the two terms only have fluxes $N_H$ and $N_{Na}$. If only HCl is present, $N_{Na}=0$ and  $N_H=N_{Cl} = -D_{HCl}\frac{d c_{Cl}}{dx}$.  Therefore, $U^a$ will be proportional to $\frac{d \log c_{Cl}}{dx}$ and reduce to the diffusiophoretic velocity due to HCl only.  This pattern is also preserved in Eq.~\eqref{eq:3ion_base} where fluxes of OH$^-$ and Cl$^-$ ions enter the expression since Na$^+$ is the common ion. We note that Eq.~\eqref{eq:3ion} requires the values of fluxes instead of concentration gradients. If concentration gradients are more readily known or calculable than fluxes, Eq.~\eqref{eq:multiU} is a more convenient choice. 
\par{} By using Eq.~\eqref{eq:3ion} to restrict $U^a \left (\zeta^a \right) > 0$ and $U^b \left (\zeta^b \right) < 0$, we obtain the conditions required for focusing 
\begin{subequations}
\label{eq:focus}
\begin{eqnarray}
    \xi^a = \frac{8 + \tilde{\zeta^a}}{8 - \tilde{\zeta^a}} > \frac{N_H/D_H +  {N_{Na}}/{D_{Na}}}{N_{Cl}/D_{Cl}} \label{eq:focus_acid}\\
     \xi^b = \frac{8 + \tilde{\zeta^b}}{8 - \tilde{\zeta^b}} < \frac{ N_{Na}/D_{Na}}{{N_{OH}}/{D_{OH}}+{N_{Cl}}/{D_{Cl}}},
\end{eqnarray}
\end{subequations}
where $\xi^{a}$ and $\xi^{b}$ are dimensionless groups, $\tilde{\zeta}_a = \frac{e \zeta_a}{k_B T}$, and $\tilde{\zeta}_b = \frac{e \zeta_b}{k_B T}$. For a given pH dependence of zeta potential, Eq.~\eqref{eq:focus} can be used to evaluate the range of salt concentrations for which the particles will focus. We do so by overlaying predictions from Eq.~\eqref{eq:focus} on Fig.~\ref{fig:focus}.
\par{} We calculate bounds set by $\xi^a$ and $\xi^b$ via Eq.~\eqref{eq:focus} through two bounds. First, for a given $\Delta \zeta$, we assume $\zeta_a$ to be the zeta potential at $x=0$ and $\zeta_b$ to the zeta potential at $x=L$. We then evaluate when the two inequalities in Eq.~\eqref{eq:focus} hold by varying $c_s$ and evaluating fluxes; see the solid lines in Fig.~\ref{fig:focus}. While this analysis accurately predicts the condition for focusing, it does differentiate between on and off-front focusing. To do so, for a given $\Delta \zeta$, we set $\zeta^{a}$ to be the zeta potential at pH$=4.5$ ad $\zeta^{b}$ to be the zeta potential at pH=$9.5$. We then reevaluate when the inequalities hold; see dashed lines in Fig.~\ref{fig:focus}. The basic idea behind this calculation assumes that the reaction front occurs across pH 4.5 to 9.5~\cite{SI}. Hence the region between dashed lines is on-front focusing whereas the region between dashed and solid lines is off-front focusing. The predictions from Eq.~\eqref{eq:focus} are thus able to correctly capture the conditions and location of focusing.

This Letter underscores the crucial physics of pH-dependent zeta potential to induce the focusing on diffusiophoretic particles on an acid-base reaction front. Looking forward, the results outlined in this manuscript can be used to explore different acid-base reaction chemistries to investigate focusing, including weak acids and bases, which will possess a weaker pH-dependence of zeta potential and thus a narrower range of focusing. More broadly, while this work focused on a one-dimensional setup, there are exciting possibilities to extend this result to  T-junctions \cite{tj1,tj2}  and 2-dimensional and porous geometries \cite{bannarjee,por1,por2,por3,por4,por5,por6}. Our work also opens up a new line of inquiry in biophysics where acid-base reactions are common and diffusiophoresis has been argued to induce pattern formation~\cite{turing} and intracellular organization~\cite{hafner2024reaction}. Finally, beyond acid-base reactions, the 3-ion extension in Eq.~\eqref{eq:3ion} to the mobility of a binary electrolyte will motivate future research with electrochemical reactions~\cite{wang2022long, wang2022visualization, jarvey2023asymmetric} and membrane systems~\cite{florea2014long} where multiple ions are common. 
\section*{Acknowledgements}
The authors would like to thank Ritu R. Raj for his insightful input into this project, and Carolyn Kohlmeier for organizing the undergraduate thesis program. The authors also thank the UROP, DLA, and YSSRP programs at the University of Colorado Boulder for financial support. A.G. thanks the NSF CAREER program (CBET-2238412) for financial support.

\bibliography{apssamp}

\begin{thebibliography}{58}%
\makeatletter
\providecommand \@ifxundefined [1]{%
 \@ifx{#1\undefined}
}%
\providecommand \@ifnum [1]{%
 \ifnum #1\expandafter \@firstoftwo
 \else \expandafter \@secondoftwo
 \fi
}%
\providecommand \@ifx [1]{%
 \ifx #1\expandafter \@firstoftwo
 \else \expandafter \@secondoftwo
 \fi
}%
\providecommand \natexlab [1]{#1}%
\providecommand \enquote  [1]{``#1''}%
\providecommand \bibnamefont  [1]{#1}%
\providecommand \bibfnamefont [1]{#1}%
\providecommand \citenamefont [1]{#1}%
\providecommand \href@noop [0]{\@secondoftwo}%
\providecommand \href [0]{\begingroup \@sanitize@url \@href}%
\providecommand \@href[1]{\@@startlink{#1}\@@href}%
\providecommand \@@href[1]{\endgroup#1\@@endlink}%
\providecommand \@sanitize@url [0]{\catcode `\\12\catcode `\$12\catcode
  `\&12\catcode `\#12\catcode `\^12\catcode `\_12\catcode `\%12\relax}%
\providecommand \@@startlink[1]{}%
\providecommand \@@endlink[0]{}%
\providecommand \url  [0]{\begingroup\@sanitize@url \@url }%
\providecommand \@url [1]{\endgroup\@href {#1}{\urlprefix }}%
\providecommand \urlprefix  [0]{URL }%
\providecommand \Eprint [0]{\href }%
\providecommand \doibase [0]{https://doi.org/}%
\providecommand \selectlanguage [0]{\@gobble}%
\providecommand \bibinfo  [0]{\@secondoftwo}%
\providecommand \bibfield  [0]{\@secondoftwo}%
\providecommand \translation [1]{[#1]}%
\providecommand \BibitemOpen [0]{}%
\providecommand \bibitemStop [0]{}%
\providecommand \bibitemNoStop [0]{.\EOS\space}%
\providecommand \EOS [0]{\spacefactor3000\relax}%
\providecommand \BibitemShut  [1]{\csname bibitem#1\endcsname}%
\let\auto@bib@innerbib\@empty
\bibitem [{\citenamefont {Palacci}\ \emph {et~al.}(2013)\citenamefont
  {Palacci}, \citenamefont {Sacanna}, \citenamefont {Steinberg}, \citenamefont
  {Pine},\ and\ \citenamefont {Chaikin}}]{crystals}%
  \BibitemOpen
  \bibfield  {author} {\bibinfo {author} {\bibfnamefont {J.}~\bibnamefont
  {Palacci}}, \bibinfo {author} {\bibfnamefont {S.}~\bibnamefont {Sacanna}},
  \bibinfo {author} {\bibfnamefont {A.~P.}\ \bibnamefont {Steinberg}}, \bibinfo
  {author} {\bibfnamefont {D.~J.}\ \bibnamefont {Pine}},\ and\ \bibinfo
  {author} {\bibfnamefont {P.~M.}\ \bibnamefont {Chaikin}},\ }\bibfield
  {title} {\bibinfo {title} {Living crystals of light-activated colloidal
  surfers},\ }\href@noop {} {\bibfield  {journal} {\bibinfo  {journal}
  {Science}\ }\textbf {\bibinfo {volume} {339}},\ \bibinfo {pages} {936}
  (\bibinfo {year} {2013})}\BibitemShut {NoStop}%
\bibitem [{\citenamefont {Aubret}\ \emph {et~al.}(2018)\citenamefont {Aubret},
  \citenamefont {Youssef}, \citenamefont {Sacanna},\ and\ \citenamefont
  {Palacci}}]{gear}%
  \BibitemOpen
  \bibfield  {author} {\bibinfo {author} {\bibfnamefont {A.}~\bibnamefont
  {Aubret}}, \bibinfo {author} {\bibfnamefont {M.}~\bibnamefont {Youssef}},
  \bibinfo {author} {\bibfnamefont {S.}~\bibnamefont {Sacanna}},\ and\ \bibinfo
  {author} {\bibfnamefont {J.}~\bibnamefont {Palacci}},\ }\bibfield  {title}
  {\bibinfo {title} {Targeted assembly and synchronization of self-spinning
  microgears},\ }\href {https://doi.org/10.1038/s41567-018-0227-4} {\bibfield
  {journal} {\bibinfo  {journal} {Nature Physics}\ }\textbf {\bibinfo {volume}
  {14}},\ \bibinfo {pages} {1114} (\bibinfo {year} {2018})}\BibitemShut
  {NoStop}%
\bibitem [{\citenamefont {Alessio}\ and\ \citenamefont {Gupta}(2023)}]{turing}%
  \BibitemOpen
  \bibfield  {author} {\bibinfo {author} {\bibfnamefont {B.~M.}\ \bibnamefont
  {Alessio}}\ and\ \bibinfo {author} {\bibfnamefont {A.}~\bibnamefont
  {Gupta}},\ }\bibfield  {title} {\bibinfo {title} {Diffusiophoresis-enhanced
  turing patterns},\ }\href@noop {} {\bibfield  {journal} {\bibinfo  {journal}
  {Science Advances}\ }\textbf {\bibinfo {volume} {9}},\ \bibinfo {pages}
  {eadj2457} (\bibinfo {year} {2023})}\BibitemShut {NoStop}%
\bibitem [{\citenamefont {Shin}\ \emph {et~al.}(2018)\citenamefont {Shin},
  \citenamefont {Warren},\ and\ \citenamefont {Stone}}]{surfactant}%
  \BibitemOpen
  \bibfield  {author} {\bibinfo {author} {\bibfnamefont {S.}~\bibnamefont
  {Shin}}, \bibinfo {author} {\bibfnamefont {P.~B.}\ \bibnamefont {Warren}},\
  and\ \bibinfo {author} {\bibfnamefont {H.~A.}\ \bibnamefont {Stone}},\
  }\bibfield  {title} {\bibinfo {title} {Cleaning by surfactant gradients:
  Particulate removal from porous materials and the significance of rinsing in
  laundry detergency},\ }\href
  {https://doi.org/10.1103/PhysRevApplied.9.034012} {\bibfield  {journal}
  {\bibinfo  {journal} {Physical Review Applied}\ }\textbf {\bibinfo {volume}
  {9}},\ \bibinfo {pages} {034012} (\bibinfo {year} {2018})}\BibitemShut
  {NoStop}%
\bibitem [{\citenamefont {Shin}\ \emph
  {et~al.}(2017{\natexlab{a}})\citenamefont {Shin}, \citenamefont {Shardt},
  \citenamefont {Warren},\ and\ \citenamefont {Stone}}]{membraneless}%
  \BibitemOpen
  \bibfield  {author} {\bibinfo {author} {\bibfnamefont {S.}~\bibnamefont
  {Shin}}, \bibinfo {author} {\bibfnamefont {O.}~\bibnamefont {Shardt}},
  \bibinfo {author} {\bibfnamefont {P.~B.}\ \bibnamefont {Warren}},\ and\
  \bibinfo {author} {\bibfnamefont {H.~A.}\ \bibnamefont {Stone}},\ }\bibfield
  {title} {\bibinfo {title} {Membraneless water filtration using {CO2}},\
  }\href@noop {} {\bibfield  {journal} {\bibinfo  {journal} {Nature
  Communications}\ }\textbf {\bibinfo {volume} {8}},\ \bibinfo {pages} {15181}
  (\bibinfo {year} {2017}{\natexlab{a}})}\BibitemShut {NoStop}%
\bibitem [{\citenamefont {Ab{\'e}cassis}\ \emph {et~al.}(2008)\citenamefont
  {Ab{\'e}cassis}, \citenamefont {Cottin-Bizonne}, \citenamefont {Ybert},
  \citenamefont {Ajdari},\ and\ \citenamefont {Bocquet}}]{boost}%
  \BibitemOpen
  \bibfield  {author} {\bibinfo {author} {\bibfnamefont {B.}~\bibnamefont
  {Ab{\'e}cassis}}, \bibinfo {author} {\bibfnamefont {C.}~\bibnamefont
  {Cottin-Bizonne}}, \bibinfo {author} {\bibfnamefont {C.}~\bibnamefont
  {Ybert}}, \bibinfo {author} {\bibfnamefont {A.}~\bibnamefont {Ajdari}},\ and\
  \bibinfo {author} {\bibfnamefont {L.}~\bibnamefont {Bocquet}},\ }\bibfield
  {title} {\bibinfo {title} {Boosting migration of large particles by solute
  contrasts},\ }\href@noop {} {\bibfield  {journal} {\bibinfo  {journal}
  {Nature Materials}\ }\textbf {\bibinfo {volume} {7}},\ \bibinfo {pages} {785}
  (\bibinfo {year} {2008})}\BibitemShut {NoStop}%
\bibitem [{\citenamefont {Prieve}\ \emph {et~al.}(1984)\citenamefont {Prieve},
  \citenamefont {Anderson}, \citenamefont {Ebel},\ and\ \citenamefont
  {Lowell}}]{ec1}%
  \BibitemOpen
  \bibfield  {author} {\bibinfo {author} {\bibfnamefont {D.}~\bibnamefont
  {Prieve}}, \bibinfo {author} {\bibfnamefont {J.}~\bibnamefont {Anderson}},
  \bibinfo {author} {\bibfnamefont {J.}~\bibnamefont {Ebel}},\ and\ \bibinfo
  {author} {\bibfnamefont {M.}~\bibnamefont {Lowell}},\ }\bibfield  {title}
  {\bibinfo {title} {Motion of a particle generated by chemical gradients.
  {P}art 2. electrolytes},\ }\href@noop {} {\bibfield  {journal} {\bibinfo
  {journal} {Journal of Fluid Mechanics}\ }\textbf {\bibinfo {volume} {148}},\
  \bibinfo {pages} {247} (\bibinfo {year} {1984})}\BibitemShut {NoStop}%
\bibitem [{\citenamefont {Anderson}(1989)}]{ec2}%
  \BibitemOpen
  \bibfield  {author} {\bibinfo {author} {\bibfnamefont {J.~L.}\ \bibnamefont
  {Anderson}},\ }\bibfield  {title} {\bibinfo {title} {Colloid transport by
  interfacial forces},\ }\href@noop {} {\bibfield  {journal} {\bibinfo
  {journal} {Annual Review of Fluid Mechanics}\ }\textbf {\bibinfo {volume}
  {21}},\ \bibinfo {pages} {61} (\bibinfo {year} {1989})}\BibitemShut {NoStop}%
\bibitem [{\citenamefont {Keh}\ and\ \citenamefont {Wei}(2000)}]{ec3}%
  \BibitemOpen
  \bibfield  {author} {\bibinfo {author} {\bibfnamefont {H.~J.}\ \bibnamefont
  {Keh}}\ and\ \bibinfo {author} {\bibfnamefont {Y.~K.}\ \bibnamefont {Wei}},\
  }\bibfield  {title} {\bibinfo {title} {Diffusiophoretic mobility of spherical
  particles at low potential and arbitrary double-layer thickness},\
  }\href@noop {} {\bibfield  {journal} {\bibinfo  {journal} {Langmuir}\
  }\textbf {\bibinfo {volume} {16}},\ \bibinfo {pages} {5289} (\bibinfo {year}
  {2000})}\BibitemShut {NoStop}%
\bibitem [{\citenamefont {Velegol}\ \emph {et~al.}(2016)\citenamefont
  {Velegol}, \citenamefont {Garg}, \citenamefont {Guha}, \citenamefont {Kar},\
  and\ \citenamefont {Kumar}}]{ec4}%
  \BibitemOpen
  \bibfield  {author} {\bibinfo {author} {\bibfnamefont {D.}~\bibnamefont
  {Velegol}}, \bibinfo {author} {\bibfnamefont {A.}~\bibnamefont {Garg}},
  \bibinfo {author} {\bibfnamefont {R.}~\bibnamefont {Guha}}, \bibinfo {author}
  {\bibfnamefont {A.}~\bibnamefont {Kar}},\ and\ \bibinfo {author}
  {\bibfnamefont {M.}~\bibnamefont {Kumar}},\ }\bibfield  {title} {\bibinfo
  {title} {Origins of concentration gradients for diffusiophoresis},\
  }\href@noop {} {\bibfield  {journal} {\bibinfo  {journal} {Soft Matter}\
  }\textbf {\bibinfo {volume} {12}},\ \bibinfo {pages} {4686} (\bibinfo {year}
  {2016})}\BibitemShut {NoStop}%
\bibitem [{\citenamefont {Ganguly}\ \emph {et~al.}(2024)\citenamefont
  {Ganguly}, \citenamefont {Roychowdhury},\ and\ \citenamefont {Gupta}}]{ec5}%
  \BibitemOpen
  \bibfield  {author} {\bibinfo {author} {\bibfnamefont {A.}~\bibnamefont
  {Ganguly}}, \bibinfo {author} {\bibfnamefont {S.}~\bibnamefont
  {Roychowdhury}},\ and\ \bibinfo {author} {\bibfnamefont {A.}~\bibnamefont
  {Gupta}},\ }\bibfield  {title} {\bibinfo {title} {Unified mobility
  expressions for externally driven and self-phoretic propulsion of
  particles},\ }\href@noop {} {\bibfield  {journal} {\bibinfo  {journal}
  {Journal of Fluid Mechanics}\ }\textbf {\bibinfo {volume} {994}},\ \bibinfo
  {pages} {A2} (\bibinfo {year} {2024})}\BibitemShut {NoStop}%
\bibitem [{\citenamefont {Chiang}\ and\ \citenamefont {Velegol}(2014)}]{ec6}%
  \BibitemOpen
  \bibfield  {author} {\bibinfo {author} {\bibfnamefont {T.-Y.}\ \bibnamefont
  {Chiang}}\ and\ \bibinfo {author} {\bibfnamefont {D.}~\bibnamefont
  {Velegol}},\ }\bibfield  {title} {\bibinfo {title} {Multi-ion
  diffusiophoresis},\ }\href@noop {} {\bibfield  {journal} {\bibinfo  {journal}
  {Journal of Colloid and Interface Science}\ }\textbf {\bibinfo {volume}
  {424}},\ \bibinfo {pages} {120} (\bibinfo {year} {2014})}\BibitemShut
  {NoStop}%
\bibitem [{\citenamefont {Shim}\ \emph {et~al.}(2022)\citenamefont {Shim},
  \citenamefont {Nunes}, \citenamefont {Chen},\ and\ \citenamefont
  {Stone}}]{ec7}%
  \BibitemOpen
  \bibfield  {author} {\bibinfo {author} {\bibfnamefont {S.}~\bibnamefont
  {Shim}}, \bibinfo {author} {\bibfnamefont {J.~K.}\ \bibnamefont {Nunes}},
  \bibinfo {author} {\bibfnamefont {G.}~\bibnamefont {Chen}},\ and\ \bibinfo
  {author} {\bibfnamefont {H.~A.}\ \bibnamefont {Stone}},\ }\bibfield  {title}
  {\bibinfo {title} {Diffusiophoresis in the presence of a ph gradient},\
  }\href@noop {} {\bibfield  {journal} {\bibinfo  {journal} {Physical Review
  Fluids}\ }\textbf {\bibinfo {volume} {7}},\ \bibinfo {pages} {110513}
  (\bibinfo {year} {2022})}\BibitemShut {NoStop}%
\bibitem [{\citenamefont {Ganguly}\ \emph {et~al.}(2023)\citenamefont
  {Ganguly}, \citenamefont {Alessio},\ and\ \citenamefont {Gupta}}]{ec8}%
  \BibitemOpen
  \bibfield  {author} {\bibinfo {author} {\bibfnamefont {A.}~\bibnamefont
  {Ganguly}}, \bibinfo {author} {\bibfnamefont {B.~M.}\ \bibnamefont
  {Alessio}},\ and\ \bibinfo {author} {\bibfnamefont {A.}~\bibnamefont
  {Gupta}},\ }\bibfield  {title} {\bibinfo {title} {Diffusiophoresis: {A} novel
  transport mechanism-fundamentals, applications, and future opportunities},\
  }\href@noop {} {\bibfield  {journal} {\bibinfo  {journal} {Frontiers in
  Sensors}\ }\textbf {\bibinfo {volume} {4}},\ \bibinfo {pages} {1322906}
  (\bibinfo {year} {2023})}\BibitemShut {NoStop}%
\bibitem [{\citenamefont {Shim}(2022)}]{ec9}%
  \BibitemOpen
  \bibfield  {author} {\bibinfo {author} {\bibfnamefont {S.}~\bibnamefont
  {Shim}},\ }\bibfield  {title} {\bibinfo {title} {Diffusiophoresis,
  diffusioosmosis, and microfluidics: {S}urface-flow-driven phenomena in the
  presence of flow},\ }\href@noop {} {\bibfield  {journal} {\bibinfo  {journal}
  {Chemical Reviews}\ }\textbf {\bibinfo {volume} {122}},\ \bibinfo {pages}
  {6986} (\bibinfo {year} {2022})}\BibitemShut {NoStop}%
\bibitem [{\citenamefont {Ault}\ and\ \citenamefont {Shin}(2024)}]{ec10}%
  \BibitemOpen
  \bibfield  {author} {\bibinfo {author} {\bibfnamefont {J.~T.}\ \bibnamefont
  {Ault}}\ and\ \bibinfo {author} {\bibfnamefont {S.}~\bibnamefont {Shin}},\
  }\bibfield  {title} {\bibinfo {title} {Physicochemical hydrodynamics of
  particle diffusiophoresis driven by chemical gradients},\ }\href@noop {}
  {\bibfield  {journal} {\bibinfo  {journal} {Annual Review of Fluid
  Mechanics}\ }\textbf {\bibinfo {volume} {57}} (\bibinfo {year}
  {2024})}\BibitemShut {NoStop}%
\bibitem [{\citenamefont {Keh}(2016)}]{ec11}%
  \BibitemOpen
  \bibfield  {author} {\bibinfo {author} {\bibfnamefont {H.~J.}\ \bibnamefont
  {Keh}},\ }\bibfield  {title} {\bibinfo {title} {Diffusiophoresis of charged
  particles and diffusioosmosis of electrolyte solutions},\ }\href@noop {}
  {\bibfield  {journal} {\bibinfo  {journal} {Current Opinion in Colloid \&
  Interface Science}\ }\textbf {\bibinfo {volume} {24}},\ \bibinfo {pages} {13}
  (\bibinfo {year} {2016})}\BibitemShut {NoStop}%
\bibitem [{\citenamefont {Kar}\ \emph {et~al.}(2015)\citenamefont {Kar},
  \citenamefont {Chiang}, \citenamefont {Ortiz~Rivera}, \citenamefont {Sen},\
  and\ \citenamefont {Velegol}}]{dep1}%
  \BibitemOpen
  \bibfield  {author} {\bibinfo {author} {\bibfnamefont {A.}~\bibnamefont
  {Kar}}, \bibinfo {author} {\bibfnamefont {T.-Y.}\ \bibnamefont {Chiang}},
  \bibinfo {author} {\bibfnamefont {I.}~\bibnamefont {Ortiz~Rivera}}, \bibinfo
  {author} {\bibfnamefont {A.}~\bibnamefont {Sen}},\ and\ \bibinfo {author}
  {\bibfnamefont {D.}~\bibnamefont {Velegol}},\ }\bibfield  {title} {\bibinfo
  {title} {Enhanced transport into and out of dead-end pores},\ }\href@noop {}
  {\bibfield  {journal} {\bibinfo  {journal} {ACS Nano}\ }\textbf {\bibinfo
  {volume} {9}},\ \bibinfo {pages} {746} (\bibinfo {year} {2015})}\BibitemShut
  {NoStop}%
\bibitem [{\citenamefont {Alessio}\ \emph {et~al.}(2021)\citenamefont
  {Alessio}, \citenamefont {Shim}, \citenamefont {Mintah}, \citenamefont
  {Gupta},\ and\ \citenamefont {Stone}}]{dep2}%
  \BibitemOpen
  \bibfield  {author} {\bibinfo {author} {\bibfnamefont {B.~M.}\ \bibnamefont
  {Alessio}}, \bibinfo {author} {\bibfnamefont {S.}~\bibnamefont {Shim}},
  \bibinfo {author} {\bibfnamefont {E.}~\bibnamefont {Mintah}}, \bibinfo
  {author} {\bibfnamefont {A.}~\bibnamefont {Gupta}},\ and\ \bibinfo {author}
  {\bibfnamefont {H.~A.}\ \bibnamefont {Stone}},\ }\bibfield  {title} {\bibinfo
  {title} {Diffusiophoresis and diffusioosmosis in tandem: Two-dimensional
  particle motion in the presence of multiple electrolytes},\ }\href@noop {}
  {\bibfield  {journal} {\bibinfo  {journal} {Physical Review Fluids}\ }\textbf
  {\bibinfo {volume} {6}},\ \bibinfo {pages} {054201} (\bibinfo {year}
  {2021})}\BibitemShut {NoStop}%
\bibitem [{\citenamefont {Alessio}\ \emph {et~al.}(2022)\citenamefont
  {Alessio}, \citenamefont {Shim}, \citenamefont {Gupta},\ and\ \citenamefont
  {Stone}}]{dep3}%
  \BibitemOpen
  \bibfield  {author} {\bibinfo {author} {\bibfnamefont {B.~M.}\ \bibnamefont
  {Alessio}}, \bibinfo {author} {\bibfnamefont {S.}~\bibnamefont {Shim}},
  \bibinfo {author} {\bibfnamefont {A.}~\bibnamefont {Gupta}},\ and\ \bibinfo
  {author} {\bibfnamefont {H.~A.}\ \bibnamefont {Stone}},\ }\bibfield  {title}
  {\bibinfo {title} {Diffusioosmosis-driven dispersion of colloids: {A}
  {T}aylor dispersion analysis with experimental validation},\ }\href@noop {}
  {\bibfield  {journal} {\bibinfo  {journal} {Journal of Fluid Mechanics}\
  }\textbf {\bibinfo {volume} {942}},\ \bibinfo {pages} {A23} (\bibinfo {year}
  {2022})}\BibitemShut {NoStop}%
\bibitem [{\citenamefont {Ault}\ \emph {et~al.}(2017)\citenamefont {Ault},
  \citenamefont {Warren}, \citenamefont {Shin},\ and\ \citenamefont
  {Stone}}]{dep4}%
  \BibitemOpen
  \bibfield  {author} {\bibinfo {author} {\bibfnamefont {J.~T.}\ \bibnamefont
  {Ault}}, \bibinfo {author} {\bibfnamefont {P.~B.}\ \bibnamefont {Warren}},
  \bibinfo {author} {\bibfnamefont {S.}~\bibnamefont {Shin}},\ and\ \bibinfo
  {author} {\bibfnamefont {H.~A.}\ \bibnamefont {Stone}},\ }\bibfield  {title}
  {\bibinfo {title} {Diffusiophoresis in one-dimensional solute gradients},\
  }\href@noop {} {\bibfield  {journal} {\bibinfo  {journal} {Soft Matter}\
  }\textbf {\bibinfo {volume} {13}},\ \bibinfo {pages} {9015} (\bibinfo {year}
  {2017})}\BibitemShut {NoStop}%
\bibitem [{\citenamefont {Wilson}\ \emph {et~al.}(2020)\citenamefont {Wilson},
  \citenamefont {Shim}, \citenamefont {Yu}, \citenamefont {Gupta},\ and\
  \citenamefont {Stone}}]{dep5}%
  \BibitemOpen
  \bibfield  {author} {\bibinfo {author} {\bibfnamefont {J.~L.}\ \bibnamefont
  {Wilson}}, \bibinfo {author} {\bibfnamefont {S.}~\bibnamefont {Shim}},
  \bibinfo {author} {\bibfnamefont {Y.~E.}\ \bibnamefont {Yu}}, \bibinfo
  {author} {\bibfnamefont {A.}~\bibnamefont {Gupta}},\ and\ \bibinfo {author}
  {\bibfnamefont {H.~A.}\ \bibnamefont {Stone}},\ }\bibfield  {title} {\bibinfo
  {title} {Diffusiophoresis in multivalent electrolytes},\ }\href@noop {}
  {\bibfield  {journal} {\bibinfo  {journal} {Langmuir}\ }\textbf {\bibinfo
  {volume} {36}},\ \bibinfo {pages} {7014} (\bibinfo {year}
  {2020})}\BibitemShut {NoStop}%
\bibitem [{\citenamefont {Gupta}\ \emph {et~al.}(2020)\citenamefont {Gupta},
  \citenamefont {Shim},\ and\ \citenamefont {Stone}}]{dep6}%
  \BibitemOpen
  \bibfield  {author} {\bibinfo {author} {\bibfnamefont {A.}~\bibnamefont
  {Gupta}}, \bibinfo {author} {\bibfnamefont {S.}~\bibnamefont {Shim}},\ and\
  \bibinfo {author} {\bibfnamefont {H.~A.}\ \bibnamefont {Stone}},\ }\bibfield
  {title} {\bibinfo {title} {Diffusiophoresis: from dilute to concentrated
  electrolytes},\ }\href@noop {} {\bibfield  {journal} {\bibinfo  {journal}
  {Soft Matter}\ }\textbf {\bibinfo {volume} {16}},\ \bibinfo {pages} {6975}
  (\bibinfo {year} {2020})}\BibitemShut {NoStop}%
\bibitem [{\citenamefont {Singh}\ \emph {et~al.}(2020)\citenamefont {Singh},
  \citenamefont {Vladisavljevi{\'c}}, \citenamefont {Nadal}, \citenamefont
  {Cottin-Bizonne}, \citenamefont {Pirat},\ and\ \citenamefont
  {Bolognesi}}]{dep7}%
  \BibitemOpen
  \bibfield  {author} {\bibinfo {author} {\bibfnamefont {N.}~\bibnamefont
  {Singh}}, \bibinfo {author} {\bibfnamefont {G.~T.}\ \bibnamefont
  {Vladisavljevi{\'c}}}, \bibinfo {author} {\bibfnamefont {F.}~\bibnamefont
  {Nadal}}, \bibinfo {author} {\bibfnamefont {C.}~\bibnamefont
  {Cottin-Bizonne}}, \bibinfo {author} {\bibfnamefont {C.}~\bibnamefont
  {Pirat}},\ and\ \bibinfo {author} {\bibfnamefont {G.}~\bibnamefont
  {Bolognesi}},\ }\bibfield  {title} {\bibinfo {title} {Reversible trapping of
  colloids in microgrooved channels via diffusiophoresis under steady-state
  solute gradients},\ }\href@noop {} {\bibfield  {journal} {\bibinfo  {journal}
  {Physical Review Letters}\ }\textbf {\bibinfo {volume} {125}},\ \bibinfo
  {pages} {248002} (\bibinfo {year} {2020})}\BibitemShut {NoStop}%
\bibitem [{\citenamefont {Singh}\ \emph {et~al.}(2022)\citenamefont {Singh},
  \citenamefont {Vladisavljevic}, \citenamefont {Nadal}, \citenamefont
  {Cottin-Bizonne}, \citenamefont {Pirat},\ and\ \citenamefont
  {Bolognesi}}]{dep8}%
  \BibitemOpen
  \bibfield  {author} {\bibinfo {author} {\bibfnamefont {N.}~\bibnamefont
  {Singh}}, \bibinfo {author} {\bibfnamefont {G.~T.}\ \bibnamefont
  {Vladisavljevic}}, \bibinfo {author} {\bibfnamefont {F.}~\bibnamefont
  {Nadal}}, \bibinfo {author} {\bibfnamefont {C.}~\bibnamefont
  {Cottin-Bizonne}}, \bibinfo {author} {\bibfnamefont {C.}~\bibnamefont
  {Pirat}},\ and\ \bibinfo {author} {\bibfnamefont {G.}~\bibnamefont
  {Bolognesi}},\ }\bibfield  {title} {\bibinfo {title} {Enhanced accumulation
  of colloidal particles in microgrooved channels via diffusiophoresis and
  steady-state electrolyte flows},\ }\href@noop {} {\bibfield  {journal}
  {\bibinfo  {journal} {Langmuir}\ }\textbf {\bibinfo {volume} {38}},\ \bibinfo
  {pages} {14053} (\bibinfo {year} {2022})}\BibitemShut {NoStop}%
\bibitem [{\citenamefont {Shin}\ \emph {et~al.}(2016)\citenamefont {Shin},
  \citenamefont {Um}, \citenamefont {Sabass}, \citenamefont {Ault},
  \citenamefont {Rahimi}, \citenamefont {Warren},\ and\ \citenamefont
  {Stone}}]{dep9}%
  \BibitemOpen
  \bibfield  {author} {\bibinfo {author} {\bibfnamefont {S.}~\bibnamefont
  {Shin}}, \bibinfo {author} {\bibfnamefont {E.}~\bibnamefont {Um}}, \bibinfo
  {author} {\bibfnamefont {B.}~\bibnamefont {Sabass}}, \bibinfo {author}
  {\bibfnamefont {J.~T.}\ \bibnamefont {Ault}}, \bibinfo {author}
  {\bibfnamefont {M.}~\bibnamefont {Rahimi}}, \bibinfo {author} {\bibfnamefont
  {P.~B.}\ \bibnamefont {Warren}},\ and\ \bibinfo {author} {\bibfnamefont
  {H.~A.}\ \bibnamefont {Stone}},\ }\bibfield  {title} {\bibinfo {title}
  {Size-dependent control of colloid transport via solute gradients in dead-end
  channels},\ }\href@noop {} {\bibfield  {journal} {\bibinfo  {journal}
  {Proceedings of the National Academy of Sciences}\ }\textbf {\bibinfo
  {volume} {113}},\ \bibinfo {pages} {257} (\bibinfo {year}
  {2016})}\BibitemShut {NoStop}%
\bibitem [{\citenamefont {Shin}\ \emph
  {et~al.}(2017{\natexlab{b}})\citenamefont {Shin}, \citenamefont {Ault},
  \citenamefont {Feng}, \citenamefont {Warren},\ and\ \citenamefont
  {Stone}}]{dep10}%
  \BibitemOpen
  \bibfield  {author} {\bibinfo {author} {\bibfnamefont {S.}~\bibnamefont
  {Shin}}, \bibinfo {author} {\bibfnamefont {J.~T.}\ \bibnamefont {Ault}},
  \bibinfo {author} {\bibfnamefont {J.}~\bibnamefont {Feng}}, \bibinfo {author}
  {\bibfnamefont {P.~B.}\ \bibnamefont {Warren}},\ and\ \bibinfo {author}
  {\bibfnamefont {H.~A.}\ \bibnamefont {Stone}},\ }\bibfield  {title} {\bibinfo
  {title} {Low-cost zeta potentiometry using solute gradients},\ }\href@noop {}
  {\bibfield  {journal} {\bibinfo  {journal} {Advanced Materials}\ }\textbf
  {\bibinfo {volume} {29}},\ \bibinfo {pages} {1701516} (\bibinfo {year}
  {2017}{\natexlab{b}})}\BibitemShut {NoStop}%
\bibitem [{\citenamefont {Akdeniz}\ \emph {et~al.}(2023)\citenamefont
  {Akdeniz}, \citenamefont {Wood},\ and\ \citenamefont {Lammertink}}]{dep11}%
  \BibitemOpen
  \bibfield  {author} {\bibinfo {author} {\bibfnamefont {B.}~\bibnamefont
  {Akdeniz}}, \bibinfo {author} {\bibfnamefont {J.~A.}\ \bibnamefont {Wood}},\
  and\ \bibinfo {author} {\bibfnamefont {R.~G.}\ \bibnamefont {Lammertink}},\
  }\bibfield  {title} {\bibinfo {title} {Diffusiophoresis and diffusio-osmosis
  into a dead-end channel: Role of the concentration-dependence of zeta
  potential},\ }\href@noop {} {\bibfield  {journal} {\bibinfo  {journal}
  {Langmuir}\ }\textbf {\bibinfo {volume} {39}},\ \bibinfo {pages} {2322}
  (\bibinfo {year} {2023})}\BibitemShut {NoStop}%
\bibitem [{\citenamefont {Shin}\ \emph
  {et~al.}(2017{\natexlab{c}})\citenamefont {Shin}, \citenamefont {Ault},
  \citenamefont {Warren},\ and\ \citenamefont {Stone}}]{tj1}%
  \BibitemOpen
  \bibfield  {author} {\bibinfo {author} {\bibfnamefont {S.}~\bibnamefont
  {Shin}}, \bibinfo {author} {\bibfnamefont {J.~T.}\ \bibnamefont {Ault}},
  \bibinfo {author} {\bibfnamefont {P.~B.}\ \bibnamefont {Warren}},\ and\
  \bibinfo {author} {\bibfnamefont {H.~A.}\ \bibnamefont {Stone}},\ }\bibfield
  {title} {\bibinfo {title} {Accumulation of colloidal particles in flow
  junctions induced by fluid flow and diffusiophoresis},\ }\href@noop {}
  {\bibfield  {journal} {\bibinfo  {journal} {Physical Review X}\ }\textbf
  {\bibinfo {volume} {7}},\ \bibinfo {pages} {041038} (\bibinfo {year}
  {2017}{\natexlab{c}})}\BibitemShut {NoStop}%
\bibitem [{\citenamefont {Liu}\ and\ \citenamefont {Pahlavan}(2025)}]{tj2}%
  \BibitemOpen
  \bibfield  {author} {\bibinfo {author} {\bibfnamefont {H.}~\bibnamefont
  {Liu}}\ and\ \bibinfo {author} {\bibfnamefont {A.~A.}\ \bibnamefont
  {Pahlavan}},\ }\bibfield  {title} {\bibinfo {title} {Diffusioosmotic reversal
  of colloidal focusing direction in a microfluidic {T}-junction},\ }\href@noop
  {} {\bibfield  {journal} {\bibinfo  {journal} {Physical Review Letters}\
  }\textbf {\bibinfo {volume} {134}},\ \bibinfo {pages} {098201} (\bibinfo
  {year} {2025})}\BibitemShut {NoStop}%
\bibitem [{\citenamefont {Chakra}\ \emph {et~al.}(2023)\citenamefont {Chakra},
  \citenamefont {Singh}, \citenamefont {Vladisavljevic}, \citenamefont {Nadal},
  \citenamefont {Cottin-Bizonne}, \citenamefont {Pirat},\ and\ \citenamefont
  {Bolognesi}}]{mc1}%
  \BibitemOpen
  \bibfield  {author} {\bibinfo {author} {\bibfnamefont {A.}~\bibnamefont
  {Chakra}}, \bibinfo {author} {\bibfnamefont {N.}~\bibnamefont {Singh}},
  \bibinfo {author} {\bibfnamefont {G.~T.}\ \bibnamefont {Vladisavljevic}},
  \bibinfo {author} {\bibfnamefont {F.}~\bibnamefont {Nadal}}, \bibinfo
  {author} {\bibfnamefont {C.}~\bibnamefont {Cottin-Bizonne}}, \bibinfo
  {author} {\bibfnamefont {C.}~\bibnamefont {Pirat}},\ and\ \bibinfo {author}
  {\bibfnamefont {G.}~\bibnamefont {Bolognesi}},\ }\bibfield  {title} {\bibinfo
  {title} {Continuous manipulation and characterization of colloidal beads and
  liposomes via diffusiophoresis in single-and double-junction microchannels},\
  }\href@noop {} {\bibfield  {journal} {\bibinfo  {journal} {ACS Nano}\
  }\textbf {\bibinfo {volume} {17}},\ \bibinfo {pages} {14644} (\bibinfo {year}
  {2023})}\BibitemShut {NoStop}%
\bibitem [{\citenamefont {Chakra}\ \emph {et~al.}(2025)\citenamefont {Chakra},
  \citenamefont {Puijk}, \citenamefont {Vladisavljevi{\'c}}, \citenamefont
  {Cottin-Bizonne}, \citenamefont {Pirat},\ and\ \citenamefont
  {Bolognesi}}]{mc2}%
  \BibitemOpen
  \bibfield  {author} {\bibinfo {author} {\bibfnamefont {A.}~\bibnamefont
  {Chakra}}, \bibinfo {author} {\bibfnamefont {C.}~\bibnamefont {Puijk}},
  \bibinfo {author} {\bibfnamefont {G.~T.}\ \bibnamefont {Vladisavljevi{\'c}}},
  \bibinfo {author} {\bibfnamefont {C.}~\bibnamefont {Cottin-Bizonne}},
  \bibinfo {author} {\bibfnamefont {C.}~\bibnamefont {Pirat}},\ and\ \bibinfo
  {author} {\bibfnamefont {G.}~\bibnamefont {Bolognesi}},\ }\bibfield  {title}
  {\bibinfo {title} {Surface chemistry-based continuous separation of colloidal
  particles via diffusiophoresis and diffusioosmosis},\ }\href@noop {}
  {\bibfield  {journal} {\bibinfo  {journal} {Journal of Colloid and Interface
  Science}\ }\textbf {\bibinfo {volume} {693}},\ \bibinfo {pages} {137577}
  (\bibinfo {year} {2025})}\BibitemShut {NoStop}%
\bibitem [{\citenamefont {Palacci}\ \emph {et~al.}(2010)\citenamefont
  {Palacci}, \citenamefont {Ab{\'e}cassis}, \citenamefont {Cottin-Bizonne},
  \citenamefont {Ybert},\ and\ \citenamefont {Bocquet}}]{trap1}%
  \BibitemOpen
  \bibfield  {author} {\bibinfo {author} {\bibfnamefont {J.}~\bibnamefont
  {Palacci}}, \bibinfo {author} {\bibfnamefont {B.}~\bibnamefont
  {Ab{\'e}cassis}}, \bibinfo {author} {\bibfnamefont {C.}~\bibnamefont
  {Cottin-Bizonne}}, \bibinfo {author} {\bibfnamefont {C.}~\bibnamefont
  {Ybert}},\ and\ \bibinfo {author} {\bibfnamefont {L.}~\bibnamefont
  {Bocquet}},\ }\bibfield  {title} {\bibinfo {title} {Colloidal motility and
  pattern formation under rectified diffusiophoresis},\ }\href@noop {}
  {\bibfield  {journal} {\bibinfo  {journal} {Physical Review Letters}\
  }\textbf {\bibinfo {volume} {104}},\ \bibinfo {pages} {138302} (\bibinfo
  {year} {2010})}\BibitemShut {NoStop}%
\bibitem [{\citenamefont {Palacci}\ \emph {et~al.}(2012)\citenamefont
  {Palacci}, \citenamefont {Cottin-Bizonne}, \citenamefont {Ybert},\ and\
  \citenamefont {Bocquet}}]{trap2}%
  \BibitemOpen
  \bibfield  {author} {\bibinfo {author} {\bibfnamefont {J.}~\bibnamefont
  {Palacci}}, \bibinfo {author} {\bibfnamefont {C.}~\bibnamefont
  {Cottin-Bizonne}}, \bibinfo {author} {\bibfnamefont {C.}~\bibnamefont
  {Ybert}},\ and\ \bibinfo {author} {\bibfnamefont {L.}~\bibnamefont
  {Bocquet}},\ }\bibfield  {title} {\bibinfo {title} {Osmotic traps for
  colloids and macromolecules based on logarithmic sensing in salt taxis},\
  }\href@noop {} {\bibfield  {journal} {\bibinfo  {journal} {Soft Matter}\
  }\textbf {\bibinfo {volume} {8}},\ \bibinfo {pages} {980} (\bibinfo {year}
  {2012})}\BibitemShut {NoStop}%
\bibitem [{\citenamefont {Banerjee}\ and\ \citenamefont
  {Squires}(2019)}]{bannarjee}%
  \BibitemOpen
  \bibfield  {author} {\bibinfo {author} {\bibfnamefont {A.}~\bibnamefont
  {Banerjee}}\ and\ \bibinfo {author} {\bibfnamefont {T.~M.}\ \bibnamefont
  {Squires}},\ }\bibfield  {title} {\bibinfo {title} {Long-range, selective,
  on-demand suspension interactions: Combining and triggering soluto-inertial
  beacons},\ }\href@noop {} {\bibfield  {journal} {\bibinfo  {journal} {Science
  Advances}\ }\textbf {\bibinfo {volume} {5}},\ \bibinfo {pages} {eaax1893}
  (\bibinfo {year} {2019})}\BibitemShut {NoStop}%
\bibitem [{\citenamefont {Raj}\ \emph {et~al.}(2023)\citenamefont {Raj},
  \citenamefont {Shields},\ and\ \citenamefont {Gupta}}]{por1}%
  \BibitemOpen
  \bibfield  {author} {\bibinfo {author} {\bibfnamefont {R.~R.}\ \bibnamefont
  {Raj}}, \bibinfo {author} {\bibfnamefont {C.~W.}\ \bibnamefont {Shields}},\
  and\ \bibinfo {author} {\bibfnamefont {A.}~\bibnamefont {Gupta}},\ }\bibfield
   {title} {\bibinfo {title} {Two-dimensional diffusiophoretic colloidal
  banding: {O}ptimizing the spatial and temporal design of solute sinks and
  sources},\ }\href@noop {} {\bibfield  {journal} {\bibinfo  {journal} {Soft
  Matter}\ }\textbf {\bibinfo {volume} {19}},\ \bibinfo {pages} {892} (\bibinfo
  {year} {2023})}\BibitemShut {NoStop}%
\bibitem [{\citenamefont {Tan}\ \emph {et~al.}(2021)\citenamefont {Tan},
  \citenamefont {Banerjee}, \citenamefont {Shi}, \citenamefont {Tang},
  \citenamefont {Abdel-Fattah},\ and\ \citenamefont {Squires}}]{por2}%
  \BibitemOpen
  \bibfield  {author} {\bibinfo {author} {\bibfnamefont {H.}~\bibnamefont
  {Tan}}, \bibinfo {author} {\bibfnamefont {A.}~\bibnamefont {Banerjee}},
  \bibinfo {author} {\bibfnamefont {N.}~\bibnamefont {Shi}}, \bibinfo {author}
  {\bibfnamefont {X.}~\bibnamefont {Tang}}, \bibinfo {author} {\bibfnamefont
  {A.}~\bibnamefont {Abdel-Fattah}},\ and\ \bibinfo {author} {\bibfnamefont
  {T.~M.}\ \bibnamefont {Squires}},\ }\bibfield  {title} {\bibinfo {title} {A
  two-step strategy for delivering particles to targets hidden within
  microfabricated porous media},\ }\href@noop {} {\bibfield  {journal}
  {\bibinfo  {journal} {Science Advances}\ }\textbf {\bibinfo {volume} {7}},\
  \bibinfo {pages} {eabh0638} (\bibinfo {year} {2021})}\BibitemShut {NoStop}%
\bibitem [{\citenamefont {Jotkar}\ \emph {et~al.}(2024)\citenamefont {Jotkar},
  \citenamefont {Ben-Noah}, \citenamefont {Hidalgo},\ and\ \citenamefont
  {Dentz}}]{por3}%
  \BibitemOpen
  \bibfield  {author} {\bibinfo {author} {\bibfnamefont {M.}~\bibnamefont
  {Jotkar}}, \bibinfo {author} {\bibfnamefont {I.}~\bibnamefont {Ben-Noah}},
  \bibinfo {author} {\bibfnamefont {J.~J.}\ \bibnamefont {Hidalgo}},\ and\
  \bibinfo {author} {\bibfnamefont {M.}~\bibnamefont {Dentz}},\ }\bibfield
  {title} {\bibinfo {title} {Diffusiophoresis of colloids in
  partially-saturated porous media},\ }\href@noop {} {\bibfield  {journal}
  {\bibinfo  {journal} {Advances in Water Resources}\ }\textbf {\bibinfo
  {volume} {193}},\ \bibinfo {pages} {104828} (\bibinfo {year}
  {2024})}\BibitemShut {NoStop}%
\bibitem [{\citenamefont {Alipour}\ \emph {et~al.}(2024)\citenamefont
  {Alipour}, \citenamefont {Li}, \citenamefont {Liu},\ and\ \citenamefont
  {Pahlavan}}]{por4}%
  \BibitemOpen
  \bibfield  {author} {\bibinfo {author} {\bibfnamefont {M.}~\bibnamefont
  {Alipour}}, \bibinfo {author} {\bibfnamefont {Y.}~\bibnamefont {Li}},
  \bibinfo {author} {\bibfnamefont {H.}~\bibnamefont {Liu}},\ and\ \bibinfo
  {author} {\bibfnamefont {A.~A.}\ \bibnamefont {Pahlavan}},\ }\bibfield
  {title} {\bibinfo {title} {Diffusiophoretic transport of colloids in porous
  media},\ }\href@noop {} {\bibfield  {journal} {\bibinfo  {journal} {arXiv
  preprint arXiv:2411.14712}\ } (\bibinfo {year} {2024})}\BibitemShut {NoStop}%
\bibitem [{\citenamefont {Warren}(2020)}]{por5}%
  \BibitemOpen
  \bibfield  {author} {\bibinfo {author} {\bibfnamefont {P.~B.}\ \bibnamefont
  {Warren}},\ }\bibfield  {title} {\bibinfo {title} {Non-faradaic electric
  currents in the nernst-planck equations and nonlocal diffusiophoresis of
  suspended colloids in crossed salt gradients},\ }\href@noop {} {\bibfield
  {journal} {\bibinfo  {journal} {Physical Review Letters}\ }\textbf {\bibinfo
  {volume} {124}},\ \bibinfo {pages} {248004} (\bibinfo {year}
  {2020})}\BibitemShut {NoStop}%
\bibitem [{\citenamefont {Williams}\ \emph {et~al.}(2024)\citenamefont
  {Williams}, \citenamefont {Warren}, \citenamefont {Sear},\ and\ \citenamefont
  {Keddie}}]{por6}%
  \BibitemOpen
  \bibfield  {author} {\bibinfo {author} {\bibfnamefont {I.}~\bibnamefont
  {Williams}}, \bibinfo {author} {\bibfnamefont {P.~B.}\ \bibnamefont
  {Warren}}, \bibinfo {author} {\bibfnamefont {R.~P.}\ \bibnamefont {Sear}},\
  and\ \bibinfo {author} {\bibfnamefont {J.~L.}\ \bibnamefont {Keddie}},\
  }\bibfield  {title} {\bibinfo {title} {Colloidal diffusiophoresis in crossed
  electrolyte gradients: Experimental demonstration of an
  “action-at-a-distance” effect predicted by the {N}ernst-{P}lanck
  equations},\ }\href@noop {} {\bibfield  {journal} {\bibinfo  {journal}
  {Physical Review Fluids}\ }\textbf {\bibinfo {volume} {9}},\ \bibinfo {pages}
  {014201} (\bibinfo {year} {2024})}\BibitemShut {NoStop}%
\bibitem [{\citenamefont {Sambamoorthy}\ and\ \citenamefont
  {Chu}(2025)}]{sambamoorthy2025diffusiophoresis}%
  \BibitemOpen
  \bibfield  {author} {\bibinfo {author} {\bibfnamefont {S.}~\bibnamefont
  {Sambamoorthy}}\ and\ \bibinfo {author} {\bibfnamefont {H.~C.}\ \bibnamefont
  {Chu}},\ }\bibfield  {title} {\bibinfo {title} {Diffusiophoresis in porous
  media saturated with a mixture of electrolytes},\ }\href@noop {} {\bibfield
  {journal} {\bibinfo  {journal} {Nanoscale Advances}\ }\textbf {\bibinfo
  {volume} {7}},\ \bibinfo {pages} {2057} (\bibinfo {year} {2025})}\BibitemShut
  {NoStop}%
\bibitem [{\citenamefont {Sambamoorthy}\ and\ \citenamefont
  {Chu}(2023)}]{sambamoorthy2023diffusiophoresis}%
  \BibitemOpen
  \bibfield  {author} {\bibinfo {author} {\bibfnamefont {S.}~\bibnamefont
  {Sambamoorthy}}\ and\ \bibinfo {author} {\bibfnamefont {H.~C.}\ \bibnamefont
  {Chu}},\ }\bibfield  {title} {\bibinfo {title} {Diffusiophoresis of a
  spherical particle in porous media},\ }\href@noop {} {\bibfield  {journal}
  {\bibinfo  {journal} {Soft Matter}\ }\textbf {\bibinfo {volume} {19}},\
  \bibinfo {pages} {1131} (\bibinfo {year} {2023})}\BibitemShut {NoStop}%
\bibitem [{\citenamefont {Shi}\ \emph {et~al.}(2016)\citenamefont {Shi},
  \citenamefont {Nery-Azevedo}, \citenamefont {Abdel-Fattah},\ and\
  \citenamefont {Squires}}]{main}%
  \BibitemOpen
  \bibfield  {author} {\bibinfo {author} {\bibfnamefont {N.}~\bibnamefont
  {Shi}}, \bibinfo {author} {\bibfnamefont {R.}~\bibnamefont {Nery-Azevedo}},
  \bibinfo {author} {\bibfnamefont {A.~I.}\ \bibnamefont {Abdel-Fattah}},\ and\
  \bibinfo {author} {\bibfnamefont {T.~M.}\ \bibnamefont {Squires}},\
  }\bibfield  {title} {\bibinfo {title} {Diffusiophoretic focusing of suspended
  colloids},\ }\href {https://doi.org/10.1103/PhysRevLett.117.258001}
  {\bibfield  {journal} {\bibinfo  {journal} {Physical Review Letters}\
  }\textbf {\bibinfo {volume} {117}},\ \bibinfo {pages} {258001} (\bibinfo
  {year} {2016})}\BibitemShut {NoStop}%
\bibitem [{\citenamefont {M{\"o}ller}\ \emph {et~al.}(2017)\citenamefont
  {M{\"o}ller}, \citenamefont {Kriegel}, \citenamefont {Kie{\ss}},
  \citenamefont {Sojo},\ and\ \citenamefont {Braun}}]{moller2017steep}%
  \BibitemOpen
  \bibfield  {author} {\bibinfo {author} {\bibfnamefont {F.~M.}\ \bibnamefont
  {M{\"o}ller}}, \bibinfo {author} {\bibfnamefont {F.}~\bibnamefont {Kriegel}},
  \bibinfo {author} {\bibfnamefont {M.}~\bibnamefont {Kie{\ss}}}, \bibinfo
  {author} {\bibfnamefont {V.}~\bibnamefont {Sojo}},\ and\ \bibinfo {author}
  {\bibfnamefont {D.}~\bibnamefont {Braun}},\ }\bibfield  {title} {\bibinfo
  {title} {Steep ph gradients and directed colloid transport in a microfluidic
  alkaline hydrothermal pore},\ }\href@noop {} {\bibfield  {journal} {\bibinfo
  {journal} {Angewandte Chemie International Edition}\ }\textbf {\bibinfo
  {volume} {56}},\ \bibinfo {pages} {2340} (\bibinfo {year}
  {2017})}\BibitemShut {NoStop}%
\bibitem [{\citenamefont {Seo}\ \emph {et~al.}(2020)\citenamefont {Seo},
  \citenamefont {Park}, \citenamefont {Lee}, \citenamefont {Lee},\ and\
  \citenamefont {Kim}}]{seo2020continuous}%
  \BibitemOpen
  \bibfield  {author} {\bibinfo {author} {\bibfnamefont {M.}~\bibnamefont
  {Seo}}, \bibinfo {author} {\bibfnamefont {S.}~\bibnamefont {Park}}, \bibinfo
  {author} {\bibfnamefont {D.}~\bibnamefont {Lee}}, \bibinfo {author}
  {\bibfnamefont {H.}~\bibnamefont {Lee}},\ and\ \bibinfo {author}
  {\bibfnamefont {S.~J.}\ \bibnamefont {Kim}},\ }\bibfield  {title} {\bibinfo
  {title} {Continuous and spontaneous nanoparticle separation by
  diffusiophoresis},\ }\href@noop {} {\bibfield  {journal} {\bibinfo  {journal}
  {Lab on a Chip}\ }\textbf {\bibinfo {volume} {20}},\ \bibinfo {pages} {4118}
  (\bibinfo {year} {2020})}\BibitemShut {NoStop}%
\bibitem [{\citenamefont {Lee}\ \emph {et~al.}(2018)\citenamefont {Lee},
  \citenamefont {Kim}, \citenamefont {Yang}, \citenamefont {Seo},\ and\
  \citenamefont {Kim}}]{lee2018diffusiophoretic}%
  \BibitemOpen
  \bibfield  {author} {\bibinfo {author} {\bibfnamefont {H.}~\bibnamefont
  {Lee}}, \bibinfo {author} {\bibfnamefont {J.}~\bibnamefont {Kim}}, \bibinfo
  {author} {\bibfnamefont {J.}~\bibnamefont {Yang}}, \bibinfo {author}
  {\bibfnamefont {S.~W.}\ \bibnamefont {Seo}},\ and\ \bibinfo {author}
  {\bibfnamefont {S.~J.}\ \bibnamefont {Kim}},\ }\bibfield  {title} {\bibinfo
  {title} {Diffusiophoretic exclusion of colloidal particles for continuous
  water purification},\ }\href@noop {} {\bibfield  {journal} {\bibinfo
  {journal} {Lab on a Chip}\ }\textbf {\bibinfo {volume} {18}},\ \bibinfo
  {pages} {1713} (\bibinfo {year} {2018})}\BibitemShut {NoStop}%
\bibitem [{\citenamefont {Gupta}\ \emph
  {et~al.}(2019{\natexlab{a}})\citenamefont {Gupta}, \citenamefont
  {Rallabandi},\ and\ \citenamefont {Stone}}]{ankur}%
  \BibitemOpen
  \bibfield  {author} {\bibinfo {author} {\bibfnamefont {A.}~\bibnamefont
  {Gupta}}, \bibinfo {author} {\bibfnamefont {B.}~\bibnamefont {Rallabandi}},\
  and\ \bibinfo {author} {\bibfnamefont {H.~A.}\ \bibnamefont {Stone}},\
  }\bibfield  {title} {\bibinfo {title} {Diffusiophoretic and diffusioosmotic
  velocities for mixtures of valence-asymmetric electrolytes},\ }\href
  {https://doi.org/10.1103/PhysRevFluids.4.043702} {\bibfield  {journal}
  {\bibinfo  {journal} {Physical Review Fluids}\ }\textbf {\bibinfo {volume}
  {4}},\ \bibinfo {pages} {043702} (\bibinfo {year}
  {2019}{\natexlab{a}})}\BibitemShut {NoStop}%
\bibitem [{\citenamefont {Florea}\ \emph {et~al.}(2014)\citenamefont {Florea},
  \citenamefont {Musa}, \citenamefont {Huyghe},\ and\ \citenamefont
  {Wyss}}]{florea2014long}%
  \BibitemOpen
  \bibfield  {author} {\bibinfo {author} {\bibfnamefont {D.}~\bibnamefont
  {Florea}}, \bibinfo {author} {\bibfnamefont {S.}~\bibnamefont {Musa}},
  \bibinfo {author} {\bibfnamefont {J.~M.}\ \bibnamefont {Huyghe}},\ and\
  \bibinfo {author} {\bibfnamefont {H.~M.}\ \bibnamefont {Wyss}},\ }\bibfield
  {title} {\bibinfo {title} {Long-range repulsion of colloids driven by ion
  exchange and diffusiophoresis},\ }\href@noop {} {\bibfield  {journal}
  {\bibinfo  {journal} {Proceedings of the National Academy of Sciences}\
  }\textbf {\bibinfo {volume} {111}},\ \bibinfo {pages} {6554} (\bibinfo {year}
  {2014})}\BibitemShut {NoStop}%
\bibitem [{\citenamefont {Timmerhuis}\ and\ \citenamefont
  {Lammertink}(2022)}]{timmerhuis2022diffusiophoretic}%
  \BibitemOpen
  \bibfield  {author} {\bibinfo {author} {\bibfnamefont {N.~A.}\ \bibnamefont
  {Timmerhuis}}\ and\ \bibinfo {author} {\bibfnamefont {R.~G.}\ \bibnamefont
  {Lammertink}},\ }\bibfield  {title} {\bibinfo {title} {Diffusiophoretic
  movements of polystyrene particles in a h-shaped channel for inorganic salts,
  carboxylic acids, and organic salts},\ }\href@noop {} {\bibfield  {journal}
  {\bibinfo  {journal} {Langmuir}\ }\textbf {\bibinfo {volume} {38}},\ \bibinfo
  {pages} {12140} (\bibinfo {year} {2022})}\BibitemShut {NoStop}%
\bibitem [{\citenamefont {Gupta}\ \emph
  {et~al.}(2019{\natexlab{b}})\citenamefont {Gupta}, \citenamefont {Shim},
  \citenamefont {Issah}, \citenamefont {McKenzie},\ and\ \citenamefont
  {Stone}}]{gupta2019diffusion}%
  \BibitemOpen
  \bibfield  {author} {\bibinfo {author} {\bibfnamefont {A.}~\bibnamefont
  {Gupta}}, \bibinfo {author} {\bibfnamefont {S.}~\bibnamefont {Shim}},
  \bibinfo {author} {\bibfnamefont {L.}~\bibnamefont {Issah}}, \bibinfo
  {author} {\bibfnamefont {C.}~\bibnamefont {McKenzie}},\ and\ \bibinfo
  {author} {\bibfnamefont {H.~A.}\ \bibnamefont {Stone}},\ }\bibfield  {title}
  {\bibinfo {title} {Diffusion of multiple electrolytes cannot be treated
  independently: Model predictions with experimental validation},\ }\href@noop
  {} {\bibfield  {journal} {\bibinfo  {journal} {Soft Matter}\ }\textbf
  {\bibinfo {volume} {15}},\ \bibinfo {pages} {9965} (\bibinfo {year}
  {2019}{\natexlab{b}})}\BibitemShut {NoStop}%
\bibitem [{SI()}]{SI}%
  \BibitemOpen
  \href@noop {} {}\bibinfo {note} {See Supplemental Material at
  URL-will-be-inserted-by-publisher for details of the theoretical
  formulation.}\BibitemShut {Stop}%
\bibitem [{\citenamefont {Ninham}\ and\ \citenamefont
  {Parsegian}(1971)}]{ninham1971electrostatic}%
  \BibitemOpen
  \bibfield  {author} {\bibinfo {author} {\bibfnamefont {B.~W.}\ \bibnamefont
  {Ninham}}\ and\ \bibinfo {author} {\bibfnamefont {V.~A.}\ \bibnamefont
  {Parsegian}},\ }\bibfield  {title} {\bibinfo {title} {Electrostatic potential
  between surfaces bearing ionizable groups in ionic equilibrium with
  physiologic saline solution},\ }\href@noop {} {\bibfield  {journal} {\bibinfo
   {journal} {Journal of Theoretical Biology}\ }\textbf {\bibinfo {volume}
  {31}},\ \bibinfo {pages} {405} (\bibinfo {year} {1971})}\BibitemShut
  {NoStop}%
\bibitem [{\citenamefont {Deen}(2012)}]{deen}%
  \BibitemOpen
  \bibfield  {author} {\bibinfo {author} {\bibfnamefont {W.~M.}\ \bibnamefont
  {Deen}},\ }\href@noop {} {\emph {\bibinfo {title} {Analysis of Transport
  Phenomena}}}\ (\bibinfo  {publisher} {Oxford University Press},\ \bibinfo
  {address} {New York},\ \bibinfo {year} {2012})\BibitemShut {NoStop}%
\bibitem [{\citenamefont {Hafner}\ and\ \citenamefont
  {Muller}(2024)}]{hafner2024reaction}%
  \BibitemOpen
  \bibfield  {author} {\bibinfo {author} {\bibfnamefont {G.}~\bibnamefont
  {Hafner}}\ and\ \bibinfo {author} {\bibfnamefont {M.}~\bibnamefont
  {Muller}},\ }\bibfield  {title} {\bibinfo {title} {Reaction-driven
  diffusiophoresis of liquid condensates: Potential mechanisms for
  intracellular organization},\ }\href@noop {} {\bibfield  {journal} {\bibinfo
  {journal} {ACS Nano}\ }\textbf {\bibinfo {volume} {18}},\ \bibinfo {pages}
  {16530} (\bibinfo {year} {2024})}\BibitemShut {NoStop}%
\bibitem [{\citenamefont {Wang}\ \emph
  {et~al.}(2022{\natexlab{a}})\citenamefont {Wang}, \citenamefont {Leville},
  \citenamefont {Behdani},\ and\ \citenamefont {Batista}}]{wang2022long}%
  \BibitemOpen
  \bibfield  {author} {\bibinfo {author} {\bibfnamefont {K.}~\bibnamefont
  {Wang}}, \bibinfo {author} {\bibfnamefont {S.}~\bibnamefont {Leville}},
  \bibinfo {author} {\bibfnamefont {B.}~\bibnamefont {Behdani}},\ and\ \bibinfo
  {author} {\bibfnamefont {C.~A.~S.}\ \bibnamefont {Batista}},\ }\bibfield
  {title} {\bibinfo {title} {Long-range transport and directed assembly of
  charged colloids under aperiodic electrodiffusiophoresis},\ }\href@noop {}
  {\bibfield  {journal} {\bibinfo  {journal} {Soft Matter}\ }\textbf {\bibinfo
  {volume} {18}},\ \bibinfo {pages} {5949} (\bibinfo {year}
  {2022}{\natexlab{a}})}\BibitemShut {NoStop}%
\bibitem [{\citenamefont {Wang}\ \emph
  {et~al.}(2022{\natexlab{b}})\citenamefont {Wang}, \citenamefont {Behdani},\
  and\ \citenamefont {Silvera~Batista}}]{wang2022visualization}%
  \BibitemOpen
  \bibfield  {author} {\bibinfo {author} {\bibfnamefont {K.}~\bibnamefont
  {Wang}}, \bibinfo {author} {\bibfnamefont {B.}~\bibnamefont {Behdani}},\ and\
  \bibinfo {author} {\bibfnamefont {C.~A.}\ \bibnamefont {Silvera~Batista}},\
  }\bibfield  {title} {\bibinfo {title} {Visualization of concentration
  gradients and colloidal dynamics under electrodiffusiophoresis},\ }\href@noop
  {} {\bibfield  {journal} {\bibinfo  {journal} {Langmuir}\ }\textbf {\bibinfo
  {volume} {38}},\ \bibinfo {pages} {5663} (\bibinfo {year}
  {2022}{\natexlab{b}})}\BibitemShut {NoStop}%
\bibitem [{\citenamefont {Jarvey}\ \emph {et~al.}(2023)\citenamefont {Jarvey},
  \citenamefont {Henrique},\ and\ \citenamefont
  {Gupta}}]{jarvey2023asymmetric}%
  \BibitemOpen
  \bibfield  {author} {\bibinfo {author} {\bibfnamefont {N.}~\bibnamefont
  {Jarvey}}, \bibinfo {author} {\bibfnamefont {F.}~\bibnamefont {Henrique}},\
  and\ \bibinfo {author} {\bibfnamefont {A.}~\bibnamefont {Gupta}},\ }\bibfield
   {title} {\bibinfo {title} {Asymmetric rectified electric and concentration
  fields in multicomponent electrolytes with surface reactions},\ }\href@noop
  {} {\bibfield  {journal} {\bibinfo  {journal} {Soft Matter}\ }\textbf
  {\bibinfo {volume} {19}},\ \bibinfo {pages} {6032} (\bibinfo {year}
  {2023})}\BibitemShut {NoStop}%
\end{thebibliography}%

\end{document}